\documentclass{iaus}
\usepackage{graphicx,ulem}

\title[APOGEE] 
{The Future of Stellar Populations Studies in the Milky Way and the Local Group}

\author[Majewski]   
{Steven R. Majewski$^1$}

\affiliation{$^1$University of Virginia, P.O. Box 400325, Charlottesville, VA 22904-4325,
USA \\ email: {\tt srm4n@virginia.edu}}

\pubyear{2009}
\volume{262}  
\pagerange{1--2}
\jname{Stellar Populations - Planning for the Next Decade}
\editors{G. Bruzual \& S. Charlot, eds.}
\begin{document}

\maketitle

\newbox\grsign \setbox\grsign=\hbox{$>$} \newdimen\grdimen \grdimen=\ht\grsign
\newbox\simlessbox \newbox\simgreatbox
\setbox\simgreatbox=\hbox{\raise.5ex\hbox{$>$}\llap
   {\lower.5ex\hbox{$\sim$}}}\ht1=\grdimen\dp1=0pt
\setbox\simlessbox=\hbox{\raise.5ex\hbox{$<$}\llap
   {\lower.5ex\hbox{$\sim$}}}\ht2=\grdimen\dp2=0pt
\def\simgreater{\mathrel{\copy\simgreatbox}}
\def\simless{\mathrel{\copy\simlessbox}}
\newbox\simppropto
\setbox\simppropto=\hbox{\raise.5ex\hbox{$\sim$}\llap
   {\lower.5ex\hbox{$\propto$}}}\ht2=\grdimen\dp2=0pt
\def\simpropto{\mathrel{\copy\simppropto}}

\begin{abstract} 

The last decade has seen enormous progress in understanding the structure
of the Milky Way and neighboring galaxies via the production of large-scale
digital surveys of the sky like 2MASS and SDSS, as well as specialized, 
counterpart imaging surveys of other Local Group systems.
Apart from providing snaphots of galaxy structure,
these ``cartographic" surveys lend insights into the formation and evolution 
of galaxies when supplemented with additional data (e.g., spectroscopy, 
astrometry) and when referenced to theoretical models and simulations of 
galaxy evolution. 
These increasingly sophisticated 
simulations are making ever more specific predictions about the detailed 
chemistry and dynamics of stellar populations in galaxies.  To fully exploit,
test and constrain these theoretical ventures demands similar commitments of 
observational effort as has been plied into the previous imaging surveys to fill 
out other dimensions of parameter space with statistically significant intensity.
Fortunately the future of large-scale stellar population studies is bright 
with a number of grand projects on the horizon that collectively will
contribute a breathtaking volume of information on individual stars in 
Local Group galaxies.  These projects include: (1) additional imaging
surveys, such as Pan-STARRS, SkyMapper and LSST, which, apart from providing
deep, multicolor imaging, yield time series data useful for revealing
variable stars (including critical standard candles, like RR Lyrae variables) 
and creating large-scale, deep proper motion catalogs; (2) higher accuracy, 
space-based astrometric missions, such as Gaia
and SIM-Lite, which stand 
to provide critical, high precision dynamical data on stars in the Milky Way
and its satellites; and (3) large-scale spectroscopic surveys 
provided by RAVE, APOGEE, HERMES, LAMOST, 
and the Gaia spectrometer, 
which will yield not only enormous numbers of stellar radial velocities, but 
extremely comprehensive views of the chemistry of stellar populations.  
Meanwhile, previously dust-obscured regions of the Milky Way will continue
to be systematically exposed via large infrared surveys underway or on the way, 
such as the various GLIMPSE surveys 
from Spitzer's IRAC 
instrument, UKIDSS, APOGEE, JASMINE and WISE.

\keywords{
Galaxy: abundances, Galaxy: halo, 
Galaxy: kinematics and dynamics, Galaxy: evolution, 
Galaxy: structure, 
Local Group, dark matter, astronomical databases, surveys}
\end{abstract}

\firstsection 

\section{Introduction: Stellar Populations Studies in the ``Industrial Age"}

The topic I have been asked to review --- the future of stellar population studies 
in the Milky Way (MW) and Local Group (LG)  --- is a vast one, within which I can 
do no more than a whirlwind tour of various topics.
To make the task more manageable, my goal will be primarily to highlight some
facets of the area of {\it resolved} stellar populations that are now, and that will soon be,
addressable as stellar populations work enters the ``industrial age" ---
an age where vast databases and automated analysis of these vast databases can
be brought to bear on both newly revealed as well as age-old problems in the
field.  To this end, a major portion of this contribution will be to summarize those
large-scale surveys about which I am aware; I apologize in advance for any
surveys I have failed to include or whose parameters I have misrepresented.

\section{Galaxy Evolution in a Cold Dark Matter Universe}

Before starting this whirlwind tour of the ``industrial age" of resolved LG stellar populations
work, it is important to be mindful of the great advances being made in sister subdisciplines,
such as cosmology, large scale structure formation, dark matter and the analysis of these
matters through large numerical simulations.
For example, it is now well established through such simulations (and with growing 
corroboration by observations)
that the growth of structure in a Cold Dark Matter (CDM) universe is 
hierarchical, with small structures merging along filaments to form larger structures, like
MW-sized galaxies and LG-sized galaxy associations.

But while hierarchical merging
until late times (even presently) is now well established as a key element of CDM structure formation
models, it is useful to recall that both late infall and hierarchical formation was anticipated as a 
central aspect of MW formation studies through the stellar populations work of 
Searle \& Zinn (1978), who showed that the outer halo globular clusters exhibit a significant
age spread 
(inferred through interpretation of
horizontal
branch morphologies).  This foundational, early result demonstrated the power of stellar
population studies for unlocking the secrets of galaxy formation.  But with $\Lambda$CDM a now
firmly entrenched and generally successful theory, it is sensible to approach future stellar population
work with this context in mind, and, in part, look to stellar population studies for testing and/or refining
this widely-accepted thesis.  This is only possible because the advanced state of 
$\Lambda$CDM simulations allows them to make a rich variety of predictions about the structure, dynamics
and chemistry of CDM structures.  But checks on the theory with resolved stellar population
studies are all the more important because, while $\Lambda$CDM has enjoyed great success in 
matching observations of structures on the largest scales, several problems still remain in 
matching the theory to observations on galaxy scales.  Among these problems on small scales 
are (1) the ``missing satellites problem", where the models predict a mass spectrum of galaxy
subhalos that are strongly in disagreement with that observed around the MW (e.g., Klypin et al. 
1999); (2) the ``central
cusps problem", where $\Lambda$CDM predicts the mass density of galaxies to  have steeply
rising central cusps (e.g., Navarro et al. 1997) for which there is little observational evidence in the
MW (see, e.g., Merrifield 2005) and which is belied by the flat central density profiles of
dwarf galaxies; and (3) problems with the predicted angular momentum distributions in galaxies, 
wherein the models have difficulty making large, extended disks like that of the MW (e.g., Abadi et al. 2003).
Thus, a current focus for advancing $\Lambda$CDM theory is attempting to resolve 
problems on galaxy scales, and clearly there are potent avenues by which 
stellar populations work in the LG, MW and their satellite systems
can contribute not only to progress in understanding hierarchical 
galaxy formation and evolution, obviously, but also to fine-tuning $\Lambda$CDM theory.

\section{Astrometry:
Testing Hierarchical Formation and Late Infall}

As one example of how stellar population studies can contribute to fine-tuning
$\Lambda$CDM theory, take the notion of infall along filaments, one of the
elements of hierachical galaxy formation vividly seen in the simulations as well as in the
observed distribution of galaxies on large scales.  Is this something we can find
evidence for in the LG?  

Infall of DM subhalos along filaments should leave rather specific dynamical fingerprints in
terms of preferred shapes, orientations and coincidences of orbits of LG systems and MW satellites, accreted 
globular clusters and (halo) stars (e.g., Knebe et al. 2004).  It has long been known that most of the
``classical" MW satellites show a relatively strong planar alignment (e.g., 
Kunkel \& Demers 1976, Kunkel 1979, Lynden-Bell 1976, 1982, Majewski 1994, Fusi Pecci et al. 1995, 
Metz et al. 2007, 2009a) almost perpendicular to the Galactic plane.
This preferential spatial distribution holds up statistically even
with the inclusion of the new ``ultrafaint" satellites revealed by the Sloan Digital Sky Survey
(SDSS), and when the orientation of the SDSS ``footprint-bias" (i.e., towards the North Galactic
Cap) is accounted for (Metz et al. 2009a).
Such spatial alignments could be a signature of a dynamical association of the satellites, but must
be checked with full 3-dimensional velocity measurements for the satellites.  Through painstaking
work, the proper motions, $\mu$, of MW satellites are slowly being accumulated, though in most cases
with still only marginally significant $\mu/\epsilon_{\mu}$, at least for ground-based data 
(see summary in Metz et al. 2008).  More recently, proper motions measured using 
Hubble Space Telescope imaging of the Magellanic Clouds (``MCs'' hereafter;
Kallivayalil et al. 2006a,b, Piatek et al.
2008), other dSph satellites (Piatek et al. 2003, 2005, 2006, 2007) and globular clusters
(e.g., Bedin et al. 2006, Kalirai et al. 2007) have shown significantly reduced proper motion
uncertainties, despite very small time baselines --- with improvements to be expected
if more epochs can be accumulated with the newly-installed WFC3.

Analysis of the proper motions of these MW satellites has more or less supported the notion
of dynamical kinship of at least some satellites, at least in that a number of these systems
seem to share similar orbital poles, showing common directions of angular momentum
(Majewski et al. 1996, Palma et al. 2002, Metz et al. 2008).  The meaning of such alignments
and whether they are at odds with, or consistent with, expectations of typical satellite
configurations around MW-like galaxies in CDM models has been much debated of late.
Among the postulated theories, such dynamical correlations could reflect 
siblinghood among satellites:
(1) through the break-up of once large satellites --- e.g., a greater Magellanic or Fornax system 
(Kunkel 1979, Lynden-Bell 1982, Majewski 1994, Metz et al. 2008); 
(2) through the formation of satellites as tidal dwarfs (e.g., Kroupa et al. 2005);
(3) through the infall of satellites onto the MW in groups (Li \& Helmi 2008, D'Onghia \& Lake 2008, 
Metz et al. 2009b); or
(4) through the preferential infall of satellites along CDM filaments (Kang et al. 2005, Zentner et al. 2005).

According to the simulations, the surviving galaxies of today reflect the most recent
infall onto the MW --- of subhalos now in the MW, those in current satellites
 were the most distant subhalos at earlier times.
Earlier material to fall in came from initially 
closer matter and is now spread out among the star and star cluster
debris of the halo.  
Thus, comparing the orbits of the satellites to the orbits of halo stars is a comparison of how infall 
proceeded at these different epochs, and this translates to spatial variations in the dynamics. 
According to high-resolution numerical simulations (e.g., Diemand et al. 2005)
the outermost Galactic halo stars and globular clusters should be on radially biased orbits
while the inner regions should contain more isotropic orbits.  
The predicted kinematics for the late-infalling satellite dwarf galaxies is
different, with small anisotropies even in the outer parts.  In addition,
the net angular momentum of these components is expected to be small and
simply reflect that of the dominant dark matter component at this epoch.

To test such predictions requires measures of halo star and satellite orbital 
anisotropy; these should be done {\it in situ}  to avoid orbital
biases attendant with solar neighborhood samples.  
(To this end, it is interesting to note the much more complex inner/outer halo structure suggested by recent analysis of the SDSS SEGUE stellar sample with proper motions by Carollo et al. 2007, 
including the suggestion that the outer halo dominates the {\it low} eccentricity component of their
relatively nearby stellar sample.)
This, in turn, 
requires the measurement of not only the bulk proper motions
of the MW satellites, but the much more demanding star-by-star measures of  
{\it distant}, individual halo stars.

\begin{table}[h]
\begin{center}
\caption{Current and Future Astrometric Surveys and Pointed Instruments}
\label{tab1}
{\scriptsize
\begin{tabular}{llcl}\hline 
{\bf Catalog\footnotemark[1]} & {\bf Accuracy} & {\bf Flux.~Limit} & {\bf Stars}\\ \hline

{\it USNO} B1.0  & 200 mas & $V\sim21$ & 1 billion \\ \hline
Tycho-2    & 60 mas & $V\sim12$ & 2.5 million \\\hline
{\it UCAC2} & 20-70 mas & $R\sim16$ & 40 million \\\hline
{\it Pan-STARRS} & 30 mas & $r\sim23$ & 2 billion \\\hline
{\it LSST}  & 3 mas & $r\sim24$ & 10 billion \\\hline
Hipparcos & 1 mas & $V\sim12$ & 117,955 \\\hline
HST+WFPC2/STIS/ACS/WFC3   & 1 mas & $V\sim24$ & pointed instrument  \\\hline
J-MAPS   &  1 mas &  $V\sim15$ & 40 million \\\hline
{\it WIYN ODI}  & 0.6 mas  & $I\sim22$  & pointed instrument \\\hline
HST/FGS   & 0.2 mas & $V\sim17$ & pointed instrument  \\\hline
JASMINE   & 0.010 mas  & $z\sim14$ & 100 million \\\hline
{\it VLBA}     & 0.010 mas & $\sim$200 mJy\footnotemark[2] & pointed instrument \\\hline
Gaia    &  0.020 mas & $V\sim15$ & 1 billion \\\hline
SIM-Lite (wide angle mode) & 0.004 mas & $V\sim20$ & pointed instrument ($\sim$10,000 sources lifetime) \\ \hline
\multicolumn{4}{l}{$\dagger${Ground-based facilities in italics.      $\ddagger$Depth depends on frequency observed.}}
\end{tabular}
}
\end{center}
\end{table}

At this dawn of the ``industrial age" of stellar populations studies we can look forward to 
astrometric studies and surveys on a vast and more precise scale  --- to the realm 
of billions of stars and  
microarcsecond precisions ---
that will help us move forward on such issues.  
Table 1 gives a sampling of astrometric capabilities 
(quoted as ``mission end" positional accuracies for the surveys and ``achieved/achievable" accuracies
for the pointed instruments {\it at the magnitude limited given})
either
already underway or expected in the next decade (with table values taken from the 
project websites or published papers).  
The projects include large ground-based surveys that 
probe deeply but at lower relative precision (USNO B1.0, UCAC2 and the future Pan-STARRS 
and LSST astrometry)
as well as more focused
ground-based facilities that can obtain higher precision --- for example VLBA observations
of radio sources (such as masers) and the soon-to-be-commissioned One Degree Imager on the
WIYN telescope, with its Orthogonal Transfer Arrays capable of on-chip local corrections for 
atmosphere-induced, differential image wander.  But upcoming
space-based astrometric missions --- including J-MAPS (which will provide a new epoch of 
positions that can be matched to those from Hipparcos and Tycho), JASMINE (a mission
to do infrared astrometry of the Galactic bulge), Gaia and SIM-Lite ---
will revolutionize the field of Galactic kinematics with astrometry reaching to microarcsecond
precision (SIM-Lite 
in narrow angle mode).  
These capabilities will enable us to achieve
the scientific goals outlined above, considering
that to derive transverse velocities of 10 km s$^{-1}$ accuracy for MW satellites at $\sim$100 kpc
requires $\sim$20 $\mu$as yr$^{-1}$
proper motions.  This is within range of SIM-Lite for individual stars and 
for Gaia after the averaging of the proper motions for many $V$$\sim$$17.5$ giant stars in each satellite.  
However,  only SIM-Lite will be able to measure the proper motions of satellites like Leo I, Leo II
or CanVen I, which are beyond 200 kpc and have their brightest members at $V$$\sim$$19.5$.  
Moreover, only SIM-Lite will be able to measure precise proper motions for many of the newfound ultrafaint
dSphs, which, even though at closer distances, have a paucity of bright giant stars.
Obviously, for single stars, where one cannot take advantage of averaging, the problem is
more acute, but Gaia will give an unparalleled view of the dynamics of inner halo stars, with
SIM-Lite providing {\it in situ} measures of stars in the outer halo.

SIM-Lite, along with VLBA astrometry on the few LG galaxies with 
detectable masers (like IC10 and M33), will allow the orbits of galaxies 
within the LG and beyond to be derived.  This 
will make it possible to constrain the matter distribution over Mpc scales
and test cosmological expectations such as infall along
filaments (Shaya et al. 2009).  There have been some tantalizing suggestions of
late LG infall by the discovery of some ``hypervelocity dwarf galaxies" near M31, namely
And XII and And XIV.  Given most reasonable estimates for the M31 mass, 
these systems have radial velocities suggesting that the dwarfs
are not presently bound to M31 or even the LG (Majewski et al. 2007, Chapman et al. 2007);
M31 would need to be at least two times more massive than previously thought to keep these 
satellites bound to it.
Meanwhile, analysis of the HST-based proper motions of the MCs (see above) by
Besla et al. (2007) strongly suggests that these galaxies are making their first pass
around the MW, which implies that they are among the latest hierarchical 
accumulations of matter by the MW.

\section{Deep, Pointed Surveys of Nearby Galaxies, Part I}

Of course, the MCs are dIrr type systems, and it is interesting that their hyperbolic orbits are consistent
with the overall distributions of dwarf satellites of different types within the LG:
It is well known that there is a general density-morphology relationship within the LG, 
with dSph and dE galaxies mostly clustered around M31 and the MW, 
while most dIrr galaxies --- the MCs among the notable and nettlesome
exceptions --- are typically found in isolated regions of  the LG.  But the new MC 
orbital data suggest (1) that the MCs violate the density-morphology relationship 
by being near the MW {\it now} due to a coincidence of fortuitous timing
in the dynamical evolution of the MCs and the LG,  and (2) that the MCs may be on the first 
stages of transformation from a dIrr morphology to another class of dwarf system.  

Such evolution of the morphologies of LG dwarf galaxies should be reflected in the
star formation histories (SFHs) of these systems.  Through systematic high resolution
imaging --- especially with the Hubble Space Telescope --- combined with stellar population synthesis,
the SFHs of these LG dwarf galaxies are being painstakingly assembled
(e.g., see summary in Dolphin et al. 2005).  This work shows clear differences in 
the SFHs
among nearby dwarf galaxies,  with a clear morphology-density-SFH relation that strongly demonstrates 
how environment is a major driver in the evolution 
of small galaxies.
This may occur through the tidal shocking of more frequent gravitational encounters with the large
LG spirals that accelerates star formation by fostering bursts and leading to an earlier exhaustion
of gas, or it may relate to this gas being stripped out by ram pressure interaction with hot coronal
gas around the large systems.  

The morphology-density-SFH relation also suggests the possibility, long discussed, that dSph 
galaxies and dIrrs may be evolutionarily related.  Such a connection is made via $N$-body simulations 
(e.g., Mayer et al. 2001a,b, Klimentowski et al. 2009) that show {\it tidal stirring} to be an active
agent transforming late type morphologies --- e.g., high surface brightness dwarf disks ---  into dSph
or dE-like
systems when placed inside a MW-sized halo.  This process involves the formation and subsequent
buckling of bars in the evolving dwarf, which transforms it from a rotation to pressure-supported
system.  In this regard it is interesting that the Large MC (and possibly the Small MC) contains a bar.
If, as these models suggest, dIrrs are simply systems that 
haven't yet been dynamically stirred, 
then the LG dIrrs should be on very large LG orbits, or they may be the most recent infalling systems 
into the LG and onto the large LG spirals (as e.g., the MCs);
thus dIrrs may represent the most ``pristine" of luminous subhalos. But much work
remains to verify this picture, especially given the arguments made by, e.g., Grebel et al. (2003)
that the central surface brightness-luminosity and luminosity-metallicity distributions
of dwarf galaxies do not favor an evolutionary connection between dIrrs and dSphs, but 
rather a closer evolutionary connection between dSphs and ``transitional" type dwarfs (like the
Antlia, Phoenix and LGS3 systems) --- a connection that, nevertheless, still belies environment (i.e., tidal 
forces) driving the differential evolution of dwarf galaxies.  
(On the other hand, Brown et al. 2007 show that the dIrr Leo A
is similar to dSphs in having a large velocity dispersion and $M/L$, i.e.  $>20\pm6$ in solar units.)

Clearly, more stellar populations work --- in particular, ever deeper HST CMDs on the LG dIrrs
--- could help resolve these questions of dwarf evolution.  The ACS Nearby Galaxy Survey Treasury
(ANGST) project is a great help in this regard, taking this detailed CMD work out to 4 Mpc 
(Dalcanton et al. 2009).
With the advent of ever larger telescopes, as well as more efficient instruments on existing 10-m
class telescopes, we also have the promise in the coming decade of detailed chemistry of 
individual stars in these systems, as is now being done on the closer, LG dSphs (see \S7).
Of course, more accurate information on the orbits of these systems, 
and understanding better the overall dynamics of the LG, would be extremely helpful 
in probing the evolutionary history of dwarf galaxies, beyond the tantalizing example offered by 
the MCs (\S3).

\section{Wide Field Photometric Surveys in the Coming Decade}

Of course, another prediction of CDM models is that infalling subhalos onto MW-like systems, 
which continue to the present, will face dynamical shredding and create a web of luminous 
(and dark) streams around these systems (e.g., Bullock \& Johnston 2005).
This vision of a MW halo networked by streams 
has been hinted at for several decades
by small, pointed  pencil-beam surveys --- with the conclusion that the 
{\it entire} halo is made from accreted systems (see, e.g., Majewski 1993, 
2004, Majewski et al. 1996).  But this vision of the MW
really has been born out by deep, wide-field photometric surveys like the 
Two Micron All-Sky Survey (2MASS) and the Sloan Digital Sky Survey (SDSS)
which have enabled the construction of large portraits of the halo where many
of these streams can be seen plainly (e.g., Newberg et al. 2002, Majewski et al. 2003, 
Belokurov et al. 2006, Grillmair 2006).  Continued active mining of these data bases continues
to unveil not only new tidal debris streams (e.g., 
Rocha-Pinto et al. 2004, Sharma et al. 2009 for 2MASS, and Belokurov et al. 2007b, Grillmair 2009,
Newberg et al. 2009 for SDSS, to name just some examples), but of course
a plethora of new dSph galaxies of extremely low luminosity
(e.g., Willman et al. 2005, Zucker et al. 2006, Belokurov et al. 2007a, Irwin et al. 2007,
Grillmair 2009).  These ultrafaint dSphs tend to have very high $M/L$ ratios
(Mu\~noz et al. 2006b, Simon \& Geha 2007) that point to a common dSph
mass scale (Strigari et al. 2008), which may arise from some critical scale in the
formation of galaxies or a characteristic scale for the clustering of DM.
But a few ultrafaints show evidence for tidal disruption and dynamical instability
(e.g., Coleman et al. 2007, Carlin et al. 2009) and may represent exceptions to the 
common mass scale (Ad{\'e}n et al. 2009).

\begin{table}[h]
\begin{center}
\caption{Wide Field Photometric Surveys Past and Future}
\label{tab2}
{\scriptsize
\begin{tabular}{llllll}\hline 
{\bf Survey} & {\bf Hemisphere} & {\bf Filters} & {\bf Magnitude Limit} & {\bf Area} & {\bf Dates}\\ \hline

2MASS &  north \& south & $J,H,K_s$ & 15.8, 15.1, 14.3 & 41,253 deg${}^{2}$ & 1997-2001 \\ \hline
SDSS I/II/III &  north & $u,g,r,i,z$ & 22.0, 22.2, 22.2, 21.3, 20.5 & 10,400 deg${}^{2}$ & 2000-2009 \\ \hline
Dark Energy Survey & south & $g,r,i,z$ &   24   & 5,000 deg${}^{2}$ &  2011-2016  \\ \hline
SkyMapper & south &  $u,v,g,r,i,z$ & 22.9, 22.7, 22.9, 22.6, 22.0, 21.5 & 20,000 deg${}^{2}$ & 2009-2014 \\ \hline
Pan-STARRS & north & $g,r,i,z,y$ & 24 &  30,000 deg${}^{2}$ & 2012-2022 \\ \hline
LSST & south & $u,g,r,i,z,y$ & 24 &  20,000 deg${}^{2}$ & 2015-2025 \\ \hline

\end{tabular}
}
\end{center}
\end{table}

The astounding pace of discovery and revolution in our understanding of the structure of our MW 
that has occurred as a result of these large-scale photometric surveys can be expected to continue 
in the next decade with impending photometric surveys that will cover a large fraction of the sky
more deeply (Table 2).  Not only will these surveys allow deep searches for streams
and satellites through color-magnitude filtering for specific stellar tracers of these structures, but
their inclusion of a time series dimension will enable the search for pulsational variables, 
which are high quality standard candles that have already proven to be quite powerful in the search
for halo substructure --- e.g., with RR Lyrae stars from SDSS (e.g., Ivezi{\'c} et al. 2004, 
Watkins et al. 2009) or the QUEST project (e.g., Duffau et al. 2006, Vivas \& Zinn 2006).

The discovery and exploration of increasing numbers of
tidal streams not only tells us about the assemblage of the {\it luminous}
MW, but also the structure and size of the MW's dark halo
because the shapes and orbits of
tidal streams are extremely sensitive probes of the overall mass distribution of the host galaxy. 
The greatest sensitivity comes with full 6-D phase space information on stream stars
(again, requiring SIM-Lite-accurate proper motions for distant streams --- e.g., 
$<10$ $\mu$as yr$^{-1}$ at $V\sim18$ for $\sim100$ kpc giant stars --- but with enormously
greater numbers of stars from Gaia exploration of closer streams).  However, great 
progress is already being made with the positions and radial velocities of Sagittarius (Sgr) stream
stars.  While up to now no study of the Sgr stream has been capable of simultanously
fitting the positions {\it and} the radial velocities of the debris (see discussion of this 
problem in Law et al. 2005), it has been shown recently (Law et al. 2009) that adoption 
of a triaxial MW dark halo solves this problem.  As more phase space data are collected 
on the Sgr and other streams sampling other Galactic radii, a more accurate description
of the MW's mass distribution will evolve; this then can be compared to expectations for
mass distributions of MW-like galaxies in CDM models.

\section{Deep, Pointed Surveys of Nearby Galaxies, Part II}

With large-scale photometric surveys like those in Table 2, we can really fine tune the predictions 
of CDM models, including an understanding of the mass distributions, orbits and timescales of 
infalling subhalos.  For example, one prediction of CDM simulations (e.g., Johnston et al. 2008)
is that a typical MW-like galaxy should presently contain on average about one large, 
high surface brightness ($\Sigma < 30$ mag arcsec$^{-2}$) stream.  Of course, in the MW such
a stream exists --- the Sgr stream (Ibata et al. 1995, 
Newberg et al. 2002, Majewski et al. 2003, Belokurov et al. 2006); 
but is the MW typical?

Fortunately, new insights into two more spiral halos are coming from
large-scale mapping of the M31 and M33 spirals --- particularly from the 
Pan-Andromeda Archaeological Survey (PAndAS) survey by Ibata
and collaborators (e.g., Ibata et al. 2007) and the Spectroscopic and Photometric Landscape of 
Andromeda's Stellar Halo (SPLASH) Survey by Guhathakurta and collaborators
(e.g., Kalirai et al. 2006).  This work has revealed the Giant Southern Stream (Ibata et al. 2001), 
a halo substructure connecting with other filagree noted around the M31 disk
(e.g., Ibata et al. 2004, Fardal et al. 2006, 2008 Gilbert et al. 2007), which makes it a 
``Sgr stream-equivalent" in our spiral twin sister.  Other M31 halo
substructure has also been revealed (e.g., Chapman et al. 2008) as well as an entire
host of newly found M31 dSph satellites (e.g., Martin et al. 2006, 2009, Majewski
et al. 2007, Zucker et al. 2007, Irwin et al. 2008, McConnachie et al. 2008).
SPLASH and PAndAS include deep spectroscopic analyses of the giant stars
discernible in M31 and allow information on the chemistry of these systems to be
gathered.  With HST, these probes of M31 can reach the main sequence turnoff
and garner information on the ages of these stellar populations as well 
(e.g., Brown et al. 2006, 2009).

In the coming decade we can expect the number of well-probed halos to
expand --- our statistical sample increased by moving to more distant
systems.  Already, significant tidal streams have been identified
through deep surface brightness imaging 
of the disk
galaxies NGC 5907 (Zheng et al. 1999, Martinez-Delgado et al. 2008b), 
NGC 4013 (Martinez-Delgado et al. 2009), M94 (Trujillo et al. 2009), and other
systems (Martinez-Delgado et al. 2008a).  Study of {\it resolved} stellar halos in more distant
galaxies
will also become more commonplace, with proven work already from the ground
(Vansevicius et al. 2004) as well as with HST (Ibata et al. 2009).

But one need not go to great distances to find additional stellar halos ---
recent work has shown that the Large MC has an extended halo (Mu\~noz et al. 2006a) with 
an exponential density profile of 4.9$^{\circ}$ scalelength (Majewski et al. 2009).  
This new Large MC halo seems to show all of the complexity of other 
halos around larger galaxies, with a mix of stars of different metallicities.  It 
includes an apparently rather metal-rich component that is similar in metallicity to the 
Small MC, and that may be related to a tidal interaction between these two systems.  In 
any case, these results suggest that even halos around smaller systems like the 
Large MC can have substructure.  Other work (e.g., Harris 2007, Noel \& Gallart 2007, 
Nidever et al. 2010) shows that even the Small MC has an 
extended outer population of stars to at least a 6$^{\circ}$ radius
that may be a halo, as is expected in hierarchical models.  
Interestingly, this extended structure seems to contain stars that are relatively young
and metal rich ([Fe/H]=$-1.0$, 2 Gyr; Nidever et al., in preparation).

\section{Current and Future Spectroscopic Surveys of the Galaxy}

Different merger histories for galaxies will yield different chemical patterns
in their substucture.  As shown by Johnston et al. (2008), 
[Fe/H] is a tracer of accreted masses (because larger mass satellites typically
have higher metallicity), while [$\alpha$/Fe] more or less traces the accretion
times of satellites (because more recent mergers --- satellites and tails --- tend to have
lower [$\alpha$/Fe]).  Thus, there are strong motivations to study the chemistry
of galactic substructure, including (1) constraining parent halo accretion histories, 
(2) learning about the SFHs and chemical enrichment histories of stream 
progenitors, (3) establishing the connection between present, bound host satellites
and stars in the host halo, (4) reconstructing the chemical distribution of the original 
satellite galaxies, (5) chemically fingerprinting stars to their parent source, 
and (6) checking the chemical make-up of halos against CDM model 
predictions.

While a lot of work has been carried out to measure detailed chemical abundance
patterns in LG dwarf galaxies and interpret them in the context of chemical
evolution models (see, e.g., the summaries by Gibson 2007 and Lanfranchi et al. 2008),
detailed chemical analysis of tidal streams is still in its early stages, with most of the focus to date on 
the MW's Sgr stream, which is relatively close and contains bright stars accessible to 
echelle resolution study.  The Sgr system shows strong population gradients along
its tidal arms (Bellazzini et al. 2006, Chou et al. 2007, Monaco et al. 2007), and illustrates
a time dependence in the chemistry of stars contributed to the MW halo as well as 
the danger of assuming that the chemistry of stars left in a disrupting dwarf are representative of
what that system contributed to the halo.  Indeed, the population gradients in the 
Sgr stream are so strong that certain dynamical inferences can be made: Either 
Sgr lost mass over a small radial range in the satellite over which there was originally
an enormous metallicity gradient, or Sgr recently suffered a more catastrophic loss
of stars over a radial range with a more typical metallicity gradient (Chou et al. 2007). 
Detailed chemical analysis of the Sgr stream also allows us to reassemble a portrait
of the Sgr progenitor; Chou et al. (2009) find that this progenitor had $\alpha$ and 
s-process element abundance patterns greatly resembling those of the Large MC, 
which seems to be a reasonable prototype for the Sgr progenitor.  Thus, by
way of stellar population similarities we obtain even more evidence pointing to the evolutionary
connection between dIrr and dSph galaxies discussed in \S4.

\begin{table}[h]
\begin{center}
\caption{Current and Future Spectroscopic Surveys of the Galaxy}
\label{tab3}
{\scriptsize
\begin{tabular}{lllll}\hline 
{\bf Survey} & {\bf R} & {\bf Stars} & {\bf Wavelength} & {\bf Dates}\\ \hline
LAMOST-LR   & 2000 & 5,000,000 & 370-900 nm & 2009-2015 \\ \hline
SEGUE       & 2000 &   240,000 & 480-920 nm & 2004-2009 \\ \hline
RAVE        & 7500 & 1,000,000 & 840-875 nm & 2003-2011 \\ \hline
LAMOST-MR   & 10,000 & 100,000 &            & 2009-2015 \\ \hline
AAOmega     & 10,000 &  50,000 & 370-950 nm & 2006-     \\ \hline
Gaia        & 11,500 & 100,000,000 & 847-874 nm & 2015-2020 \\ \hline
APOGEE      & 30,000 & 100,000 & 1520-1690 nm & 2011-2014 \\ \hline
HERMES      & 30,000 & 1,200,000 & 370-950 nm & 2011-2012 \\ \hline
WINERED     & 100,000 & 1,000,000 & 900-1300 nm & TBD \\ \hline
\end{tabular}
}
\end{center}
\end{table}

Significant advances in the understanding of the chemical evolution of our galaxy
and its network of disrupted dSphs can be expected in the next decade, where
huge databases of medium to high resolution spectroscopy of Galactic stars will
be generated with a variety of instruments (Table 3).  Obviously, the accumulation of 
high quality radial velocities for many millions of stars will also shape a very detailed understanding
of the dynamics of Galactic stellar populations.  Not only will the Galactic halo
and its substructure be thoroughly probed by these upcoming surveys, but 
so too will the populations of the inner Galaxy --- including those typically hidden by 
dust obscuration --- by infrared spectroscopic surveys like APOGEE 
(see Schiavon et al., this proceedings).

\section{Probing the Inner Galaxy with Infrared Surveys}

Of course, $\Lambda$CDM surveys
show galaxies forming {\it entirely} from hierarchical merging in an ``inside-out" manner, 
so detailed studies of the inner MW are essential to unlocking the dust-concealed story of its 
early formation.  Because of this dust obscuration, explorations at infrared wavelengths hold 
the most promise for this work.
Fortunately, the coming decade brings a host of new infrared capabilities to add to the
near-infrared (NIR) imaging brought by 2MASS and other more focused surveys.  
The UKIDSS project (Lucas et al. 2008) will increase the depth of these NIR imaging probes at low 
Galactic  latitudes, and also provide second epoch positional data that can be combined with
2MASS for proper motions (e.g., Deacon et al. 2009).  A more focused effort of $z$-band
astrometry of the Galactic disk and bulge is the goal of the 
Japan Astrometry Satellite Mission for Infrared Exploration (JASMINE; Gouda et al. 2005).

The Spitzer Space Telescope, which has significantly opened up longer wavelength
photometric and spectroscopic study of the inner MW, 
will continue to provide longer wavelength imaging capability
even as it enters its more wavelength-limited, Warm Mission.  In particular, the Infrared Array Camera (IRAC)
will continue to map the Galactic midplane at 3.6 and 4.5$\mu$m through the GLIMPSE-360 
mission, extending the earlier GLIMPSE I and II projects (Churchwell et al. 2009), which have
been 
instrumental in uncovering the structure of the central Galaxy
(e.g., Benjamin et al. 2005).
Even more coverage (wavelength and area)
will be given by the Wide-Field Infrared Survey Explorer (WISE), to be launched in the autumn 
of 2009.
And APOGEE 
will provide high resolution NIR spectroscopy of stars otherwise inaccessible behind
tens of magnitudes of optical extinction. 

Despite the diminished effects of infrared studies to the presence of dust, it cannot 
be completely ignored.  Continued efforts must be made to understand the Galactic
extinction law, especially given that it is not universal either by ISM density or
by position in the Galaxy (e.g., Fitzpatrick \& Massa 2009, Zasowski et al. 2009, Gao et al. 2009).
Fortunately, the combination of NIR and MIR photometry (e.g., 2MASS/UKIDSS + GLIMPSE/WISE)
gives direct information on the reddening foreground to each star, which will make it possible in the future
to develop maps not only of the 3-D distribution of stars in the Galactic disk,
but also of the dust (Zasowski et al., in preparation).

\acknowledgements
I appreciate the assistance of Richard J. Patterson and Jeffrey L. Carlin with collecting the data for
and creating the tables in this paper.  I acknowledge support from NSF
grants AST-0607726 and AST-0807945,  
as well as support by the {\it SIM Lite} key project {\it Taking Measure of
the Milky Way} under NASA/JPL contract 1228235.



\end{document}